\documentclass{aa}
\usepackage{psfig}
\usepackage{natbib}

\bibpunct{(}{)}{;}{a}{}{,}

\begin{document}

   \thesaurus{06            
              (08.02.1;     
               08.05.3;     
               08.13.2;     
               08.23.1)}    

\title{Reconstructing the evolution of double helium white dwarfs:\\
envelope loss without spiral-in}

\author{Gijs Nelemans\inst{1}, Frank
Verbunt\inst{2}, Lev R. Yungelson\inst{1,3}  and Simon F. Portegies
Zwart\inst{4}\thanks{Hubble Fellow}}

\offprints{Gijs Nelemans}

\institute{Astronomical Institute ``Anton Pannekoek'', 
        Kruislaan 403, NL-1098 SJ Amsterdam, the Netherlands, gijsn@astro.uva.nl 
                \and 
           Astronomical Institute, Utrecht University,
           P.O.Box 80000, NL-3508 TA Utrecht, the Netherlands, verbunt@astro.uu.nl
             \and
             Institute of Astronomy of the Russian Academy of
             Sciences, 48 Pyatnitskaya Str., 109017 Moscow, Russia, lry@inasan.rssi.ru 
             \and
        Department of Astronomy Boston University,
            725 Commonwealth Avenue, Boston, MA 01581, USA, spz@komodo.bu.edu
}

\date{Received April 26, 2000; in original form February 4,
  2000/Accepted June 8, 2000}
\maketitle

\markboth{G. Nelemans et al. : Binary evolution reconstruction}{}

\begin{abstract}
  
  The unique core-mass -- radius relation for giants with degenerate
  helium cores enables us to reconstruct the evolution of three
  observed double helium white dwarfs with known masses of both
  components.  
  
  The last mass transfer phase in their evolution must have been a
  spiral-in. In the formalism proposed by \citet{web84}, we
  can constrain the efficiency of the deposition of orbital
  energy into the envelope to be $1 \la \alpha \la 6$, for an envelope
  structure parameter $\lambda=0.5$. We find that the two standard
  mass transfer types (stable mass transfer and spiral-in) are both
  unable to explain the first phase of mass transfer for these three
  binaries.
  
  We use a parametric approach to describe mass transfer in low-mass
  binaries, where both stars are of comparable mass and find that the
  orbital characteristics of the observed double helium white dwarfs
  can be well reproduced if the envelope of the primary is lost with
  $\sim$1.5 times the specific angular momentum of the initial binary.
  In this case no substantial spiral-in occurs.


\end{abstract}

\section{Introduction}

The long lasting problem that we observe many double stars which are
expected to form close pairs of white dwarfs, but yet that of the
observed white dwarfs not one seemed to have a close white dwarf
companion, was solved by the discovery of such pairs, starting with
L870-2 (= WD 0135+052) in 1988 \citep{slo88}.  In total 14 close
detached binary white dwarfs are known at present, see
Table~\ref{tab:obs}.  The fact that six of these systems
have their orbital period and the masses of both components determined
provides an opportunity to test binary evolution theory in detail.

Models for the formation of close double white dwarfs envision two
standard scenarios to produce these systems
\citep{ty81,it84b,web84,ty88,hpe95,ity97,han98}. In the first
scenario, two low-mass ($M \la 2.3\,\mbox{${\rm M}_{\sun}$}$) stars
evolve through two stages of spiral-in.  The first spiral-in will
shrink the orbit, so the second spiral-in happens in a binary with a
much smaller orbital separation.  The Roche lobe filling giant
(secondary) now has a small radius and therefore a small core. The
white dwarf that is formed last is thus less massive than its
companion, with a mass ratio $m_{\rm bright}/m_{\rm dim} \la 0.55$
\citep[see e.g.][]{stb00}.

In the second scenario, the first phase of mass transfer is stable;
the second phase of mass transfer is again a spiral-in. The evolution
of the orbit and the growth of the core during the first, slow phase
of mass transfer depend on the amount of mass and angular momentum
that is lost from the system. If the evolution in this phase is
conservative, the expected final mass ratio $m_{\rm bright}/m_{\rm
  dim} \approx 1.14 - 1.18$ \citep{ty88,stb00}.

All white dwarfs in close pairs known today have low masses ($M \la
0.5 \mbox{${\rm M}_{\sun}$}$).  Note, however, that the inaccuracy of
the mass determinations is as large as $\sim 0.05\mbox{${\rm
    M}_{\sun}$}$\ due to uncertainties in model atmospheres and
cooling curves for white dwarfs \citep[see e.g.][]{ngs99}.  These low
masses suggest they are helium white dwarfs, but it cannot be excluded
{\em a priori} that white dwarfs with masses $\ga 0.35\,\mbox{${\rm
    M}_{\sun}$}$\ are so called hybrid white dwarfs, i.e.  having
small CO cores and relatively thick ($\sim 0.1\,\mbox{${\rm
    M}_{\sun}$}$) helium envelopes \citep{it85}. For the most massive
ones $(M \ga 0.45\,\mbox{${\rm M}_{\sun}$})$, this is even the only
option, since helium white dwarfs must have a mass below 0.46
$\mbox{${\rm M}_{\sun}$}$ \citep{sgr90}. For the less massive ones the
probability to form hybrid white dwarfs is 4 -- 5 times lower than to
form helium white dwarfs \citep{nyp+00}.

\begin{table}[t]
\caption[]{Parameters of known close double white dwarfs 
with $m_{\rm WD}$\ denoting the mass of the brightest white dwarf
and $M_{\rm WD}$\ denoting its companion. For references see
\citet{mm99}$^1$ and \citet{mmm+00}.
}
\label{tab:obs}
\begin{center}
\begin{tabular}{cccc} \hline
WD & $P$(d) & $m_{\rm WD}/\mbox{${\rm M}_{\sun}$}$  & $M_{\rm WD}/\mbox{${\rm M}_{\sun}$}$  \\ \hline
0135$-$052 & 1.556  & 0.47 & 0.52\\
0136$+$768 & 1.407  & 0.44 & 0.34\\
0957$-$666 & 0.061  & 0.37 & 0.32\\
1101$+$364 & 0.145  & 0.31 & 0.36\\ 
1204$+$450 & 1.603  & 0.51 & 0.51\\ 
1704$+$481 & 0.145  & 0.39 & 0.56\\ \hline
1022$+$050 & 1.157  & 0.35 &     \\
1202$+$608 & 1.493  & 0.40 &     \\
1241$-$010 & 3.347  & 0.31 &     \\
1317$+$453 & 4.872  & 0.33 &     \\
1713$+$332 & 1.123  & 0.38 &     \\
1824$+$040 & 6.266  & 0.39 &     \\
2032$+$188 & 5.084  & 0.36 &     \\
2331$+$290 & 0.167  & 0.39 &     \\ \hline
\end{tabular}
\end{center}
$^1$ For WD 0136+768 we give the masses of components
  after \citet{bsl92}, correcting a misprint in \citet{mm99}.
\end{table}

\section{Reconstructing the binary evolution}\label{reconstruct}

Because of the unique core-mass -- radius relation for giants with
degenerate helium cores \citep{rw70}, we can reconstruct the mass
transfer phases in which helium white dwarfs are formed.  The mass of
the brightest star in WD~1101+364 (0.31 $\mbox{${\rm M}_{\sun}$}$)
indicates that it is a helium white dwarf. In WD~136+768 and
WD~0957-666 it cannot be excluded that the brightest stars are hybrid
white dwarfs.  The low mass of the dimmer companions in these three
systems indicates that those are all helium white dwarfs. The white
dwarfs in WD~0135-052 and WD~1204+450 are formally inconsistent with
being helium white dwarfs, so we will not include them in the
discussion anymore. WD 1704+481 probably consists of a helium white
dwarf and a dimmer CO white dwarf, but because CO white dwarfs cool
faster than helium white dwarfs \citep{dsb+98}, it is not clear which
of the white dwarfs was formed most recent so we cannot use this
system in our present study.

Assuming that the mass of the white dwarf is equal to the mass of the
core of the giant at the onset of the mass transfer, the radius of the
progenitor of a helium white dwarf can be calculated from the
core-mass -- radius relation given by \citet{it85}:
\begin{equation}\label{eq:Rg}
R  \approx 10^{3.5}  \; M_{\rm c}^4
\end{equation}
($R$ and $M_{\rm c}$ in solar units). This equation is in good
agreement with other equations describing this dependence for giants
\citep[e.g.][]{wrs83}.  The {\em mass of the white dwarf progenitor}
is however not known, since the above relation is independent of the
total mass of the star. However the mass of the giant must be in the
range 0.8 - 2.3 $\mbox{${\rm M}_{\sun}$}$.  For less massive stars the
main-sequence life time is larger than the age of the Galaxy. More
massive stars do not form degenerate helium cores.

For the remainder of this article we use the following notation:
$M_{\rm i}$ and $M_{\rm WD}$ indicate the initial mass of the original
primary and the mass of the white dwarf that it forms, $m_{\rm i}$ and
$m_{\rm WD}$ represent the same for the original secondary. If the
secondary accretes mass during the first phase of mass transfer, we
represent its new mass with $m_{\rm i}^\prime$. For the radii of the stars
when they become giants we use $R_{\rm g}$ and $r_{\rm g}$ for the
original primary and secondary respectively. With $P_{\rm i}, P_{\rm
  m}$ and $P$ we indicate the initial period, the period after
the first phase of mass transfer, and the current period of the binary.

\section{Last mass transfer: spiral-in}

\begin{figure}[t]
\psfig{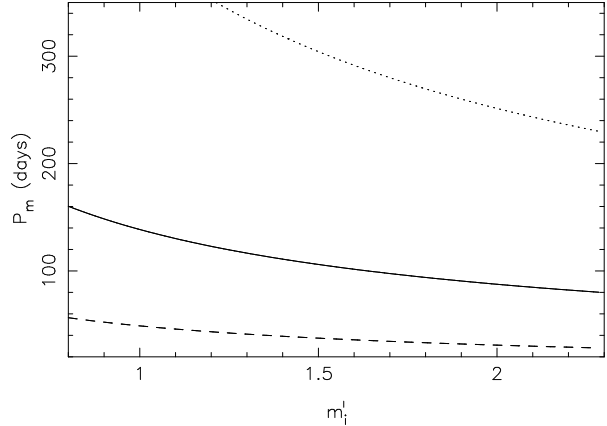}
\caption[]{Periods before the spiral-in 
phase in which the younger white dwarf was formed as function of
the mass of its giant progenitor.
Lines from bottom to top: WD 1101+346,
0957-666 and 0136+768}
\label{fig:Pm}
\end{figure}

Using Eq.~(\ref{eq:Rg}) we calculate the radii of the progenitors of
the brightest white dwarfs for the three double helium white dwarfs.
Since we know the mass of the white dwarf that orbited this giant and
may reasonably assume that it did not accrete anything during the
spiral-in phase, we can calculate the orbital separation at the onset
of the spiral-in, as function of the mass $m_{\rm i}^\prime$ of the
giant,
\begin{equation}\label{eq:am}
a_{\rm m} (m_{\rm i}^\prime) = \frac{r_{\rm g}(m_{\rm WD})}{r_{\rm
    L}(M_{\rm WD}/m_{\rm i}^\prime)},
\end{equation}
where $r_{\rm L} \equiv R_{\rm L}/a$ is the dimensionless Roche lobe
radius, given e.g. by \citet{egg83} and we assume $r_{\rm g}
= R_{\rm L}$. This is shown in Fig.~\ref{fig:Pm}, where we use Keplers
3$^{\rm rd}$ law to compute the period from the orbital separation.
Comparing the periods in Fig.~\ref{fig:Pm} with the observed periods in
Table~\ref{tab:obs}, we see that in the last mass transfer phase the
orbital separation must have reduced dramatically. This can only be
accomplished if the last mass transfer was a spiral-in.

In a spiral-in, the envelope of the giant is expelled at
the expense of the orbital energy of the binary.  Balancing the
binding energy of the envelope of the giant with the difference in orbital
energy \citep{web84} one finds
\begin{equation}\label{eq:ce}
\frac{M_{\rm WD} \; (m_{\rm i}^\prime - m_{\rm WD})}{\lambda \; r_{\rm g}}  
 = \alpha \; \left[ \frac{M_{\rm WD}\; m_{\rm WD}}{2 \;a_{\rm
       f}} -  \frac{M_{\rm WD}\; m_{\rm i}^\prime}{2
 \;a_{\rm m}} \right].
\end{equation}
The parameter $\lambda$\ depends on the structure of red giant
envelope. The usual assumption is that $\lambda = 0.5$\ \citep{khp87}.
The parameter $\alpha$\ represents the efficiency of the deposition of
orbital energy into the common envelope.  To reduce the number of
parameters, the product $\alpha \lambda$\ is treated as a single
parameter in the remainder of this article, but it should be noted
that both $\lambda$\ and $\alpha$\ will in reality be functions of the
evolutionary stage of the stars.

\begin{figure}[t]
\psfig{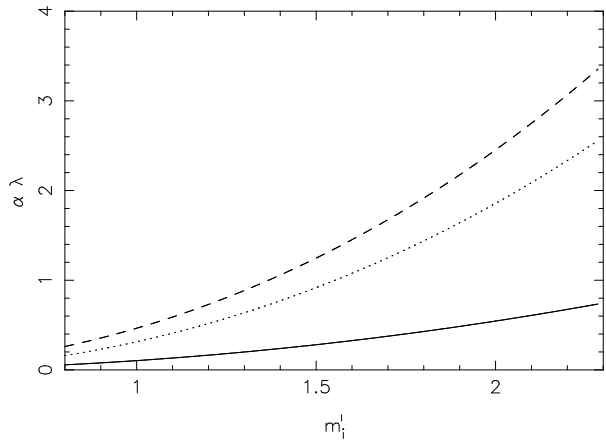}
\caption{The parameter $\alpha \lambda$ for WD~0957-666, 1101+364 and
  0136+768 assuming that the brightest component is a helium white
  dwarf and their orbital periods did not change since the end of the
  spiral-in stage. Lines from bottom to top are for WD~0957-666,
  0136+768 and 1101+364.}
\label{fig:al}
\end{figure}

Applying Eq.~(\ref{eq:ce}) and Eq.~(\ref{eq:am}), we find $\alpha
\lambda$\ as a function of $m_{\rm i}^\prime$ for the three systems
considered. We plot this in Fig.~\ref{fig:al},
where we assume that the current periods are equal to the post
spiral-in periods. This may not be the case in general since close
orbits like the those of WD 0957-666 and WD 1101+364 will decay due to
the loss of angular momentum by gravitational radiation.  However, the
estimated ages for these white dwarfs ($\sim 10^7$ yr \citep{mmb97}
and $\sim 10^9$ yr (using the cooling curves of \citet{dsb+98})
respectively) are short compared to the orbital decay time scale.

For the remaining white dwarf pairs listed in Table~\ref{tab:obs} the
mass of only one component is known. We assume here that it is the
last formed component we observe.  Low-mass white dwarfs may have
thick hydrogen envelopes which make them cool very slowly
\citep{dsb+98} and the situation in which the older white dwarf is
really observed cannot be excluded {\it a priori}. However, as we show
in an forthcoming paper \citep{nyp+00}, in the majority of binary
white dwarfs we indeed observe the youngest of the two dwarfs.
 
\begin{figure}[t]
\psfig{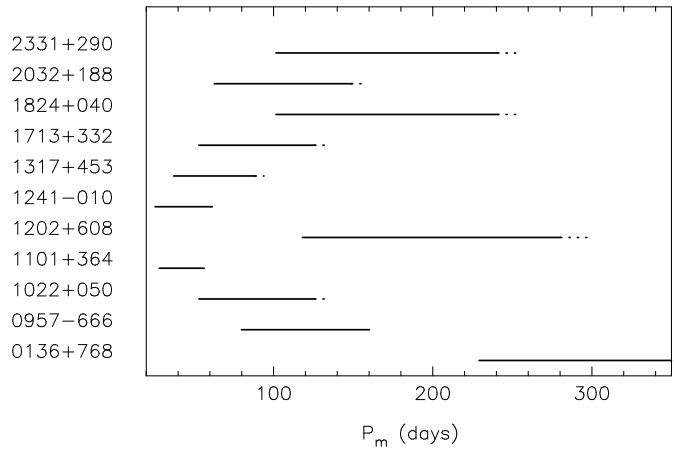}
\caption{Periods before the spiral-in for all double white
dwarfs in which the last formed is a helium white dwarf. Limits allow
for the (unknown) mass of the progenitor of this white dwarf and the
mass of the unseen companion. Solid lines are for a companion mass between
0.2 and 0.65 $\mbox{${\rm M}_{\sun}$}$. Dotted line gives the limit for a companion mass
of 0.65 - 1.4 $\mbox{${\rm M}_{\sun}$}$}
\label{fig:Pm_hist}
\end{figure}

\begin{figure}[t]
\psfig{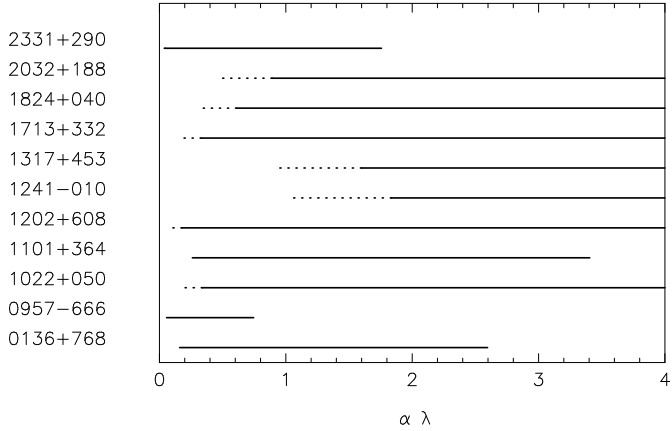}
\caption{The parameter $\alpha \lambda$ for the same cases as in Fig.~\ref{fig:Pm_hist}}
\label{fig:al_hist}
\end{figure}

From Figs.~\ref{fig:Pm} and \ref{fig:al} we see that we find a range
of $P_{\rm m}$'s and $\alpha \lambda$'s for WD 0957-666, 1101+364 and
0136+768 where the mass of the second white dwarf is known. For the
remaining systems we can also compute a range of $P_{\rm m}$'s and
$\alpha \lambda$'s by determining the ranges for all possible masses
of the unseen companion. We do know that the companion almost
certainly is another white dwarf, so the mass of this object must be
between 0.2 and 1.4\,$\mbox{${\rm M}_{\sun}$}$ and most probably even
below 0.65\,$\mbox{${\rm M}_{\sun}$}$ \citep{ity97,han98,nyp+00}.
Having also in mind that the mass of white dwarf progenitor is in the
0.8 -- 2.3 \mbox{${\rm M}_{\sun}$} range we can derive the possible
ranges of intermediate periods and $\alpha \lambda$'s for all double
white dwarfs, as is shown in Figs.~\ref{fig:Pm_hist} and
\ref{fig:al_hist}.


\noindent Two conclusions can be drawn from Figs.~\ref{fig:Pm} to ~\ref{fig:al_hist}: 
\begin{enumerate}
\item The efficiency of the energy deposition into the common
  envelopes $\alpha$\ must be high.  From model calculations of
  stellar structure we know that $ \lambda \sim 0.5 -1.0.$ If we
  assume that $\alpha$\ does not depend on the evolutionary state of
  the giant or the combination of the masses of the giant and the
  white dwarf, the parameters of all double white dwarfs with two
  observed helium components can be reproduced with the same $\alpha
  \sim 4$. The only exception is WD 0957-666 for which the efficiency
  appears to be much lower (see Sect. \ref{sec:0957co} for a different
  solution). Note that an error in the masses of the white dwarfs of
  0.05 \mbox{${\rm M}_{\sun}$}\ translates to an error in the value of
  $\alpha \lambda$\ of a factor $\sim$1.5. However even this does not
  bring the value of $\alpha$\ to 4 for WD 0957-666.
  
  Since Eq.~(\ref{eq:ce}) only considers a rough energy budget, the
  conclusion $\alpha > 1$ could simply mean that we do not
  calculate the energy accurately. It does mean that the orbital
  energy deposition into common envelope has to be highly efficient.
  It could also mean that sources other than the orbital energy
  contribute to the process of common envelope expulsion \citep{il93}.
  E.g. it is possible that the envelope is partially removed before
  Roche lobe contact by an enhanced stellar wind due to tidal
  interaction between the giant and the companion \citep{te88},
  yielding a lower value of $\alpha \lambda$.
  
\item The immediate progenitors of the known close double white dwarfs
  (i.e. the white dwarf + giant binaries) all had rather wide orbits
  (25 -- $>$500 days). This has important consequences for the
  understanding of the first phase of mass transfer.
\end{enumerate}

\section{The first mass transfer}

We compute the evolution of the binary parameters in the first mass
transfer, where we start from the initial binary and evolve it forward
according to the two standard scenario's. The resulting periods should
be equal to the intermediate periods we reconstructed in the previous
section.

\subsection{Spiral-in}

\begin{figure}[t]
\psfig{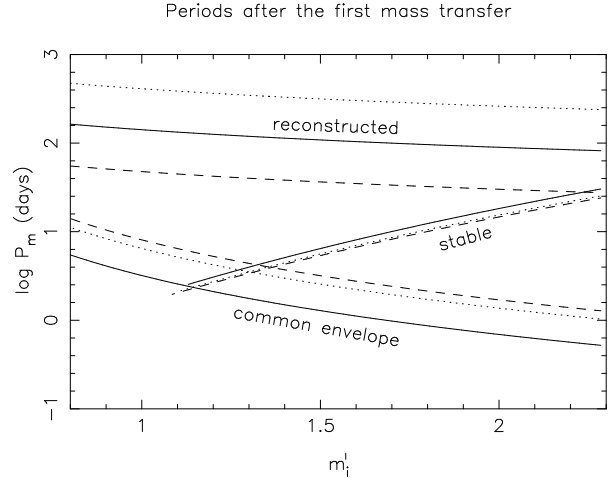}
\caption[]{Periods after the first phase of mass transfer ($P_{\rm
    m}$) as function of the mass of secondary at this time ($m_{\rm
    i}^\prime$). Top three lines are periods needed to explain the
  mass of the second formed white dwarf (see Fig.~\ref{fig:Pm}).
  Middle three lines give the maximum period would the first phase of
  mass transfer be an Algol phase, lower three ones are for the case
  the first mass transfer phase was a spiral-in.  Solid lines
  for WD 0957-666, dashed lines for WD 1101+346 and dotted lines for
  WD 0136+768.}
\label{fig:Pm_log}
\end{figure}

In the case when the first mass transfer was also a spiral-in, we can
compute the period after the spiral-in by making a (standard)
assumption that $\alpha \lambda$ is constant (i.e. here, as derived
above $\alpha \lambda = 2$). The initial separation at which the
primary fills its Roche lobe is again determined by Eq.~(\ref{eq:Rg})
and a value for $M_{\rm i}$ between 0.8 and 2.3 $\mbox{${\rm
    M}_{\sun}$}$.  Applying Eq.~(\ref{eq:ce}), we find $a_{\rm m}$
(and thus $P_{\rm m}$) as a function of $M_{\rm i}$ and $m_{\rm i}$.
The maximum period after the spiral-in is given for the case $M_{\rm
  i} = m_{\rm i}$, which we plot in Fig.~\ref{fig:Pm_log} (bottom
three lines). These periods are clearly much smaller than the periods
derived in the previous section (see Fig.~\ref{fig:Pm}). We reproduce
the latter on a logarithmic scale as the top three lines in
Fig.~\ref{fig:Pm_log}. In a recent article \citet{stb00} came to the
same conclusion, based on the observed mass ratios and also concluded
that it is very hard for binaries containing two low-mass stars to
survive two spiral-in phases.

It may be argued that $\alpha \lambda$ is different in the first mass
transfer phases, because the companion now is a main-sequence star
instead of a white dwarf. We can, just as in the case of the second
mass transfer determine the value of $\alpha \lambda$ that is required
to get the right period after the mass transfer.  {\em If we do this,
  however, we find $-15 \la \alpha \lambda \la -5$\, clearly out of the
  allowed range.  This means that a spiral-in phase in the first
  mass transfer phase for these systems is ruled out.}

\subsection{Stable mass transfer}\label{stable1st}

For the alternative scenario, to start with a short period zero-age
binary and evolve through a phase of stable mass transfer, the
zero-age orbital period should be in a narrow interval, such that the
primary fills its Roche lobe at the moment it still has a radiative or
at least shallow convective envelope.  For these stars mass transfer
in which the donor stays in thermal equilibrium is possible if the
initial mass ratio $q_{\rm i} \equiv m_{\rm i}/M_{\rm i} \ga 0.8$
\citep{tfy82}. For more extreme mass ratio's; $q_{\rm i} \ga 0.3$,
mass transfer can proceed on a timescale in between the thermal and
the dynamical timescale of the donor (e.g. trial computations for a
system with $M_{\rm i} = 2.3 \mbox{${\rm M}_{\sun}$}, m_{\rm i} = 0.8
\mbox{${\rm M}_{\sun}$}$ and $a_{\rm i} = 12.8 R_{\sun}$ show that
$\log \dot{M_{\rm i}} \la -5.5 \mbox{${\rm M}_{\sun}$}$ yr$^{-1}$; A.
Fedorova, private communication).  However for mass ratio's $q_{\rm i}
\la 0.5$, the thermal timescale of the accretor is much longer than
the mass transfer timescale and the accretor is expected to expand,
leading to a contact configuration which is unstable
\citep[e.g.][]{km77}, unless the secondary is a convective low-mass
main-sequence star \citep{web77}.

Since we want to explore the limits of the Algol scenario, we allow
all values of $q_{\rm i} > 0.3$ and calculate the possible periods
after the Algol phase.
  
We assume the mass transfer is conservative and for a given total mass
of the system $M_{\rm tot} = M_{\rm wd} + m_{\rm i}^\prime$, the
initial masses are $M_{\rm i} = M_{\rm tot}/(1 + q_{\rm
  i})$ and $m_{\rm i} = q_{\rm i} M_{\rm i}$. We consider systems in
which the primary fills its Roche lobe just before the star develops a
relatively deep convective envelope (the outer 50\% of the mass is
convective). The radii of such stars are obtained by a fit to stellar
models by \citet{mds+79}: $R_{\rm max} \approx 2.4 \; M_{\rm
  i}^{1.56}$ (solar units). Using the approximation to the Roche lobe
given by \citet{pac67} and the equation for the change in period for
conservative mass transfer we find the period after the Algol
phase is
\begin{equation}\label{eq:P_m}
P_{\rm m} \approx 1.38  \frac{M_{\rm tot}^{7.84}}{(M_{\rm wd} \;  m_{\rm
    i}^{\prime} \,)^3} \; \frac{ q_{\rm i}^3}{ (1 + q_{\rm i})^{7.84}}
\qquad \mbox{days.}
\end{equation} 
This equation has a maximum for $q_{\rm i} = 0.62$.  We compute these
maximal periods as function of $m_{\rm i}^\prime$, where of course
$M_{\rm i}$ and $m_{\rm i}$ are limited by $M_{\rm i} > 0.8
\mbox{${\rm M}_{\sun}$}$, because the primary has to evolve off the
main-sequence and $m_{\rm i}^\prime < 2.3 \mbox{${\rm M}_{\sun}$}$,
because the secondary must still develop a degenerate helium core.

The resulting periods for $q = 0.62$ are shown as the middle set
of lines in Fig.~\ref{fig:Pm_log}, where we actually used the equation
for the Roche lobe given by \citet{egg83}, which is better in
this mass ratio regime. The obtained periods are clearly not long
enough to explain of the origin of WD 0957-666, 1101+364 and 0136+768.

One could argue that the assumption of conservative evolution is not
correct.  \citet{han98}, for example, in his `best' model assumes 50\%
of the mass is lost with the specific angular momentum of the donor.
This could yield wider orbits than obtained above. Looking at his Fig.
7 one sees that he can indeed explain WD 1101+364, but not WD 0957-666
and especially not WD 0136+768, because for these systems the masses
of the last formed white dwarfs (i.e. the periods after the Algol
phase) are too large. To get higher masses one needs to lose less
angular momentum with the mass, leading to even wider orbits.
However, analysis of observed low-mass Algols, which are binaries
currently in this stage of stable mass exchange
\citep[e.g.][]{rrw74,my75,gia81,it84b,kty86,mh96a} shows that their
periods are smaller than would be expected with Eq.~(\ref{eq:P_m}).
This suggests that descendants of Algol-type systems have orbital
separations even smaller than in the case of conservative evolution,
which we assumed for Fig.~\ref{fig:Pm_log}.

{\em We conclude that with the above assumptions Algol evolution in
  the first mass transfer stage for WD 0136+768, 0957-66 and 1101+364
  is ruled out also.}

There are some other observed systems with white dwarfs which probably
could not be formed through Algol-type evolution
(Table~\ref{tab:wdms}), since white dwarf + giant binaries with
$M_{\rm WD} \la 0.25 \mbox{${\rm M}_{\sun}$}$\ end their Algol phase
with periods below $\sim$30 d. \citep{kty86}. In another system, HD
185510, with $P_{\rm orb}$=20.7 day, a giant has a hot companion which
is classified as a 0.3 $\mbox{${\rm M}_{\sun}$}$ sdB star from its
temperature and gravity \citep{fmc98}.  However, it also fits the
range of temperatures and gravities for $\sim 0.25 \mbox{${\rm
    M}_{\sun}$}$\ helium white dwarfs \citep{dsb+98} and the system
may be a viable Algol descendant. If it is an sdB star, which {\it a
  priori} is less likely, it actually matches a scenario similar to
one for WD 0957-666 rather well (see Sect. \ref{sec:0957co}).

\begin{table}[t]
\caption{Parameters of wide binaries with giant and white dwarf
  components. Masses in solar units.}
\label{tab:wdms}
\begin{center}
\begin{tabular}{lrrrr} \hline
Name            & $P$ (d) & $M$  & $m_{\rm WD}$ & ref  \\ \hline
S1040           & 42.8    & 1.5        & 0.22 & \citet{lab+97}\\ 
AY Cet          & 56.8    & 2.2        & 0.25 & \citet{sfg85}\\ \hline
\end{tabular}
\end{center}
\end{table}

\section{Unstable mass transfer revised}

Since both standard scenarios for the first phase of mass transfer
appear to be ruled out, the situation apparently is more complex.
\citet{te88} \citep[see also][]{han98} assume that due to tidal
effects of the companion, the star can lose up to 150 times more mass
in the stellar wind than without companion.  This has of course the
desired effect that the orbit widens before the mass transfer starts,
and that there is less envelope mass left to be expelled, leading to a
less dramatic spiral-in. However, if we recompute the lines for a
spiral-in in Fig.~\ref{fig:Pm_log} with reduced envelopes such that
they overlap with the reconstructed lines, we find that for WD
1101+364 we need to reduce the envelope by 70\% and for the others by
even more than 90\%.
  
In searching for a different solution we start by noting that the
original spiral-in picture \citep{pac76} considers a companion which
orbits inside the envelope of the giant.  The companion experiences
drag forces while moving in the envelope and frictional effects brake
the companion. In this process orbital energy is transformed into heat
and motion of the gas and finally into kinetic energy that causes the
envelope of the giant to be expelled.  This picture is very much based
on the situation where there is a tidal instability which causes the
decay of the orbit of the companion in systems with a high mass ratios
of the components \citep{dar08,cou73}.  In the case that the common
envelope is caused by a runaway mass transfer, the common envelope
will not look much like the equilibrium envelope of the star and
worse, the angular momentum of the orbit is so large that the common
envelope is in principle easily brought into co-rotation with the
orbit. At that moment there are no drag forces anymore.

Since for stars with deep convective envelopes mass loss on a
dynamical time scale seems, in the current state of the art in stellar
evolution modelling, inevitable, we have to assume that in the
progenitors of the observed double white dwarfs some kind of common
envelope engulfing the whole system forms. The parameters of the
observed close binary white dwarfs suggest that this envelope is
subsequently lost without much spiral-in. The energy to expel the
envelope may be supplied by the luminosity of the giant or by tidal
heating, or a combination of both. In absence of a detailed physical
description we will describe the effects of this mechanism in terms of
the angular momentum balance.

We compare the pre- and post-mass transfer binaries, under the
assumption that the envelope of the giant is lost completely from the
binary (i.e.  $m_{\rm i}^{\prime} = m_{\rm i}$), and that this mass
loss reduces the angular momentum of the system in a linear way, as
first suggested for the general case of non-conservative mass transfer
by \citet{pz67}
\begin{equation}
J_{\rm i} - J_{\rm m} = \gamma J_{\rm i} \frac{\Delta M}{M_{\rm tot}},
\end{equation}
where $J_{\rm i}$ is the angular momentum of the pre-mass transfer
binary and $M_{\rm tot}$ is the total mass of the binary.  The change
of the  orbital period as function of the initial and final
masses of the components then becomes
\begin{equation}\label{eq:orbit_change}
\frac{P_{\rm m}}{P_{\rm i}} = \left(\frac{M_{\rm WD} m_{\rm i}^{\prime}}{M_{\rm
i} m_{\rm i}}\right)^{-3} \!
\left(\frac{M_{\rm WD} + m_{\rm i}^{\prime}}{M_{\rm i} + m_{\rm i}}\right) \! \left(1 - \gamma 
\frac{\Delta M}{M_{\rm i} + m_{\rm i}}\right)^3.
\end{equation}

\begin{figure}[t]
\psfig{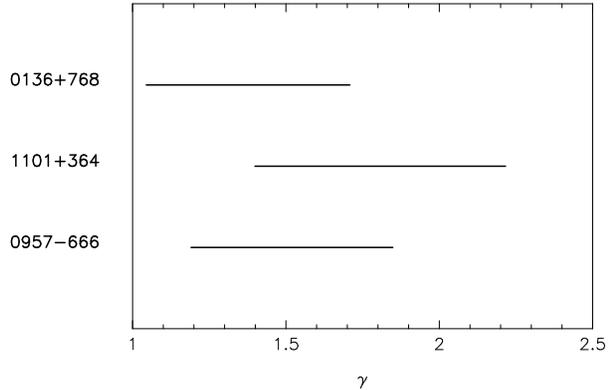}
\caption[]{Derived range of possible values of $\gamma$, for the three
double helium white dwarfs.}
\label{fig:gamma_hist}
\end{figure}

We can estimate $\gamma$ for the three double helium white dwarf
systems, just as for the case of $\alpha \lambda$.  It turns out that
all three systems are consistent with a value of $\gamma$ between
$\sim$1.4 and $\sim$1.7 (Fig.~\ref{fig:gamma_hist}).

Thus the mass transfer from a giant to a main-sequence star may in
general either be stable (in the case where the giant still has a
radiative, or at least not too deep convective envelope), unstable,
leading to a spiral-in, or a process in which the envelope is lost
without much of a spiral-in.  Which systems do and which do not
experience a spiral-in is related to the mass ratio of the components.
As can be seen from Eq.~(\ref{eq:orbit_change}) for systems with $
\gamma \; \Delta M \approx M_{\rm i} \; (1 + q)$ the periods already
become very small and the effect is essentially the same as in the
case of a spiral-in.

\section{Formation of observed systems}\label{examples}

\begin{figure*}[ht]
\begin{tabular}{c|c|c}
\begin{minipage}{0.305\textwidth}
\centerline{\psfig{figure=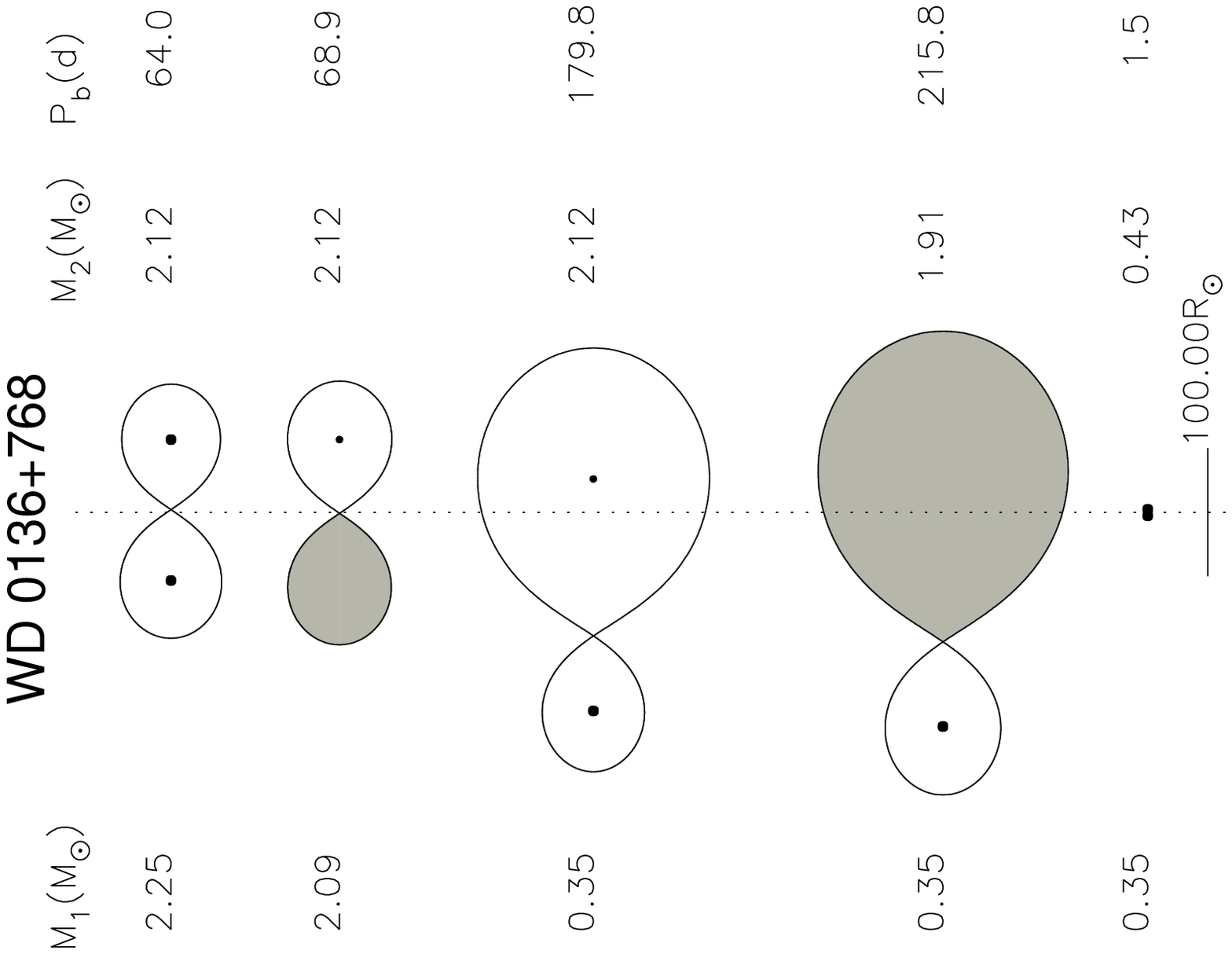,angle=-90,width=\columnwidth,bbllx=550pt,bblly=310pt,bburx=20pt,bbury=675pt,clip=}}
\end{minipage}
&
\begin{minipage}{0.345\textwidth}
\centerline{\psfig{figure=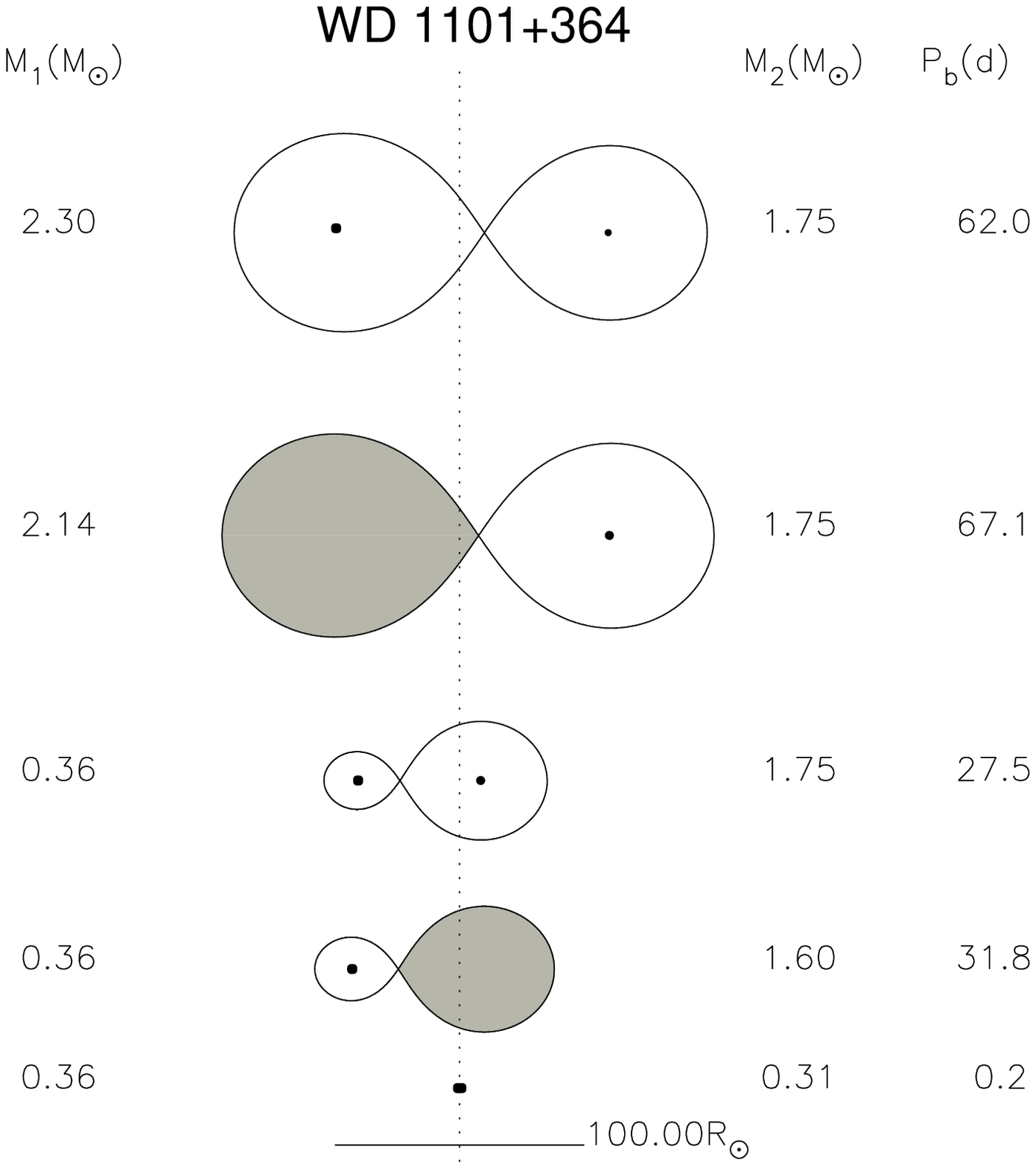,angle=-90,width=\columnwidth,bbllx=550pt,bblly=285pt,bburx=20pt,bbury=695pt,clip=}}
\end{minipage}
&
\begin{minipage}{0.30\textwidth}
\centerline{\psfig{figure=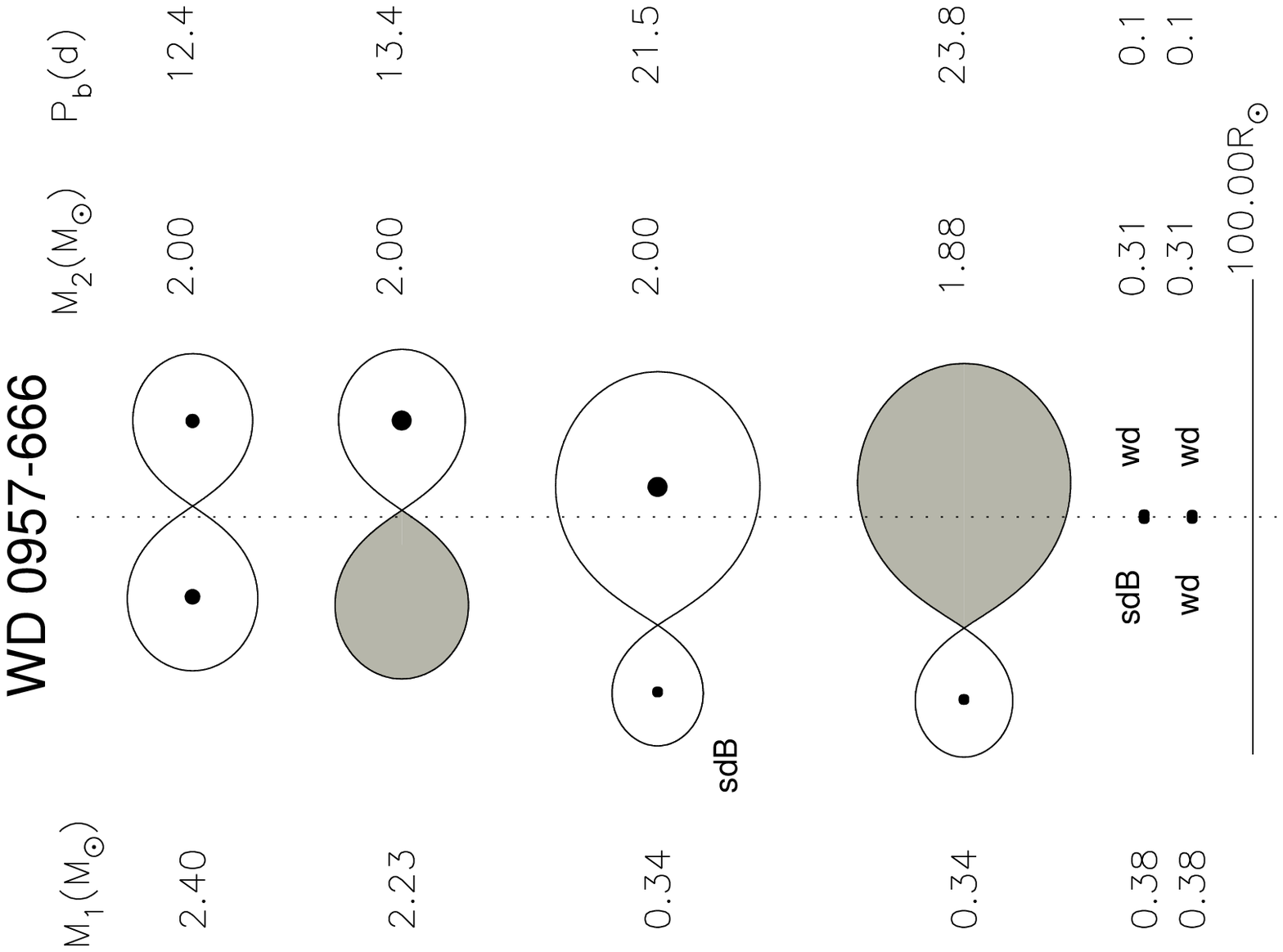,angle=-90,width=\columnwidth,bbllx=565pt,bblly=310pt,bburx=20pt,bbury=675pt,clip=}}
\end{minipage}
\end{tabular}
\caption[]{Evolutionary scenarios for the formation of WD 0136+768, WD 
  1101+364 and WD 0957-666 (left to right). In all scenario's the
  primaries lose their envelope after filling their Roche lobe,
    causing a change in the orbital period described by
    Eq.~(\ref{eq:orbit_change}).  Whether the orbit widens (WD
  0136+768 and WD 0957-666) or shrinks (WD 1101+364) depends on the
  mass ratio.  The second mass transfer always results in a spiral-in.
  For WD 0957-666 we present a scenario in which in the first mass
  transfer a helium star (sdB star) is formed which becomes a hybrid
  white dwarf only after the companion has become a helium white dwarf
  in the second phase of mass transfer.
}
\label{fig:scenarios}
\end{figure*}

From Figs.~\ref{fig:al} and \ref{fig:al_hist} it's clear that WD
0136+768, WD 0957-666 and WD 1101+364 could be formed with $\alpha
\lambda$\ below 0.8.  However from Fig.~\ref{fig:al_hist} it looks
like the extremely short-period system WD~0957-66 falls out of the
sample in the sense that the other systems, as well as systems with
unobserved companions, are compatible with a value of $\alpha\lambda
\sim 2.$ 

\subsection{Formation of helium white dwarf pairs}

In Fig.~\ref{fig:scenarios} we show evolutionary scenarios for WD
0136+768 (left) and WD 1101+364 (middle) in which they consist of two
helium white dwarfs. We included the effect of stellar winds which was
not taken into account in the preceding discussion. Therefore we use
slightly different values for $\gamma$ (1.75 and 1.85 respectively)
than the values derived in Fig.~\ref{fig:gamma_hist}.  The difference
in the initial mass ratio of the components results in dramatically
different orbital periods after the envelope of the primary is lost:
almost tripling in the first case and decrease by half in the second.
As a result, after the second mass transfer episode the second white
dwarf is the more massive member of the pair in WD~0136+768 but the
less massive in WD~1101+364.  The large difference in periods and
especially masses after the first phase of mass transfer, results in
rather different final orbital periods.

For WD~0957-666 a scenario in which both components are helium white
dwarfs can also be constructed but then one has to assume that $\alpha
\lambda$ is atypically small. However we suggest a different scenario.

\subsection{An alternative scenario for WD 0957-666: carbon-oxygen
  white dwarf with helium companion}
\label{sec:0957co}

There is another solution which allows us to explain the origin of
WD~0957-666 using $\alpha \lambda \approx 2$\ and $\gamma \approx
1.75$, like for the other two systems.

The mass of the observable white dwarf in WD~0957-666 system allows it
to be a hybrid white dwarf \citep{it85}. Such white dwarfs descend
from stars with initial mass 2.3 $\mbox{${\rm M}_{\sun}$}$ - 5
$\mbox{${\rm M}_{\sun}$}$, which fill their Roche lobes in the stage
of hydrogen burning in a shell around a non-degenerate helium core,
become hot subdwarfs in the core helium-burning stage, but don't
experience envelope expansion after the formation of a degenerate
carbon-oxygen core. Their masses are between 0.33 and 0.8 $\mbox{${\rm
    M}_{\sun}$}$. The formation of hybrid white dwarfs was considered
in the study of the population of white dwarfs by \citet[see their
scenario 3]{ty93} and all their following studies.

In a scenario shown in Fig.~\ref{fig:scenarios} (right) we start with
a system of 2.4\,\mbox{${\rm M}_{\sun}$} and 2.0\,\mbox{${\rm
    M}_{\sun}$} in a relatively close orbit $(a_0 \approx
37\,R_{\sun})$. At the instant of Roche lobe overflow the primary has
a deep convective envelope and we apply Eq.~(\ref{eq:orbit_change}) to
compute the change in the orbital period. The primary becomes a
compact helium star.  A peculiarity of low-mass helium stars is their
long life time, $\sim 1.1 \times 10^7 M^{-3.75}$ yr for $0.33 \la
M/\mbox{${\rm M}_{\sun}$} \la 0.7$\ \citep{pcw+91}, comparable to the
lifetime of their main-sequence progenitors.  As a consequence, the
initially slightly less massive secondary fills its Roche lobe and
becomes a helium white dwarf while the former primary still burns
helium in its core.  For some 250 Myr the system could be observed as
a hot subdwarf with a companion unseen due to the difference in
luminosities. After core helium exhaustion the subdwarf cools and
becomes a ``hybrid'' white dwarf.  \citet{mmb97} estimate from its
$T_{\rm eff}$\ that the cooling age of this white dwarf is only $\sim
10^7$\,yr and that the ratio of the luminosities of components is
close to 5. This is compatible with the age of about 250 Myr expected
in our scenario for the 0.31\,\mbox{${\rm M}_{\sun}$} companion.

\section{Conclusion}

We followed the binary evolution for three double helium white dwarfs
backwards and came to the following conclusions.
\begin{enumerate}
\item The last phase of mass transfer (the primary has already become
  a white dwarf and the secondary fills its Roche lobe) was a
  spiral-in, for which we can constrain $\alpha$, which describes the
  efficiency of orbital energy deposition into the common envelope, to
  lie between 1 and 6, assuming a structure parameter $\lambda = 0.5$.
  This efficiency value may be an overestimate since $\lambda$\ may
  increase towards 1 at the end of the first red giant stage.  Our
  result is in agreement with values of $\alpha \sim 4$\ found in
  population synthesis studies of low-mass X-ray binaries
  \citep{tau96}, double neutron stars \citep{py98} and double white
  dwarfs \citep{nyp+00}.
\item The parameters of all observed double helium white dwarfs may be
  reproduced with the same $\alpha \lambda \approx 2$.
\item WD 0957-666 is the only system for which $\alpha \lambda$\ 
    appears to be lower if both components are helium dwarfs. However,
  this system might have been formed with $\alpha \lambda \approx 2$\ 
  if the immediate precursor of the currently observed white
  dwarf was a non-degenerate helium star and now it is a hybrid white
  dwarf.
\item In order to explain the relatively high masses of the observed
  white dwarfs in close pairs, their direct progenitors (i.e. white
  dwarf + giant binaries) must have had relatively wide orbits
  (between 25 and $>$500 days). The standard cases of mass transfer
  (Algol evolution and spiral-in) applied to the first phase of mass
  transfer, can not explain these intermediate wide orbits.  Only if
  the masses of the observed white dwarfs are (much) lower than the
  current estimates (i.e. below 0.3 $\mbox{${\rm M}_{\sun}$}$) they
  could be formed through a phase of stable mass transfer.
\item We suggest that in the first mass transfer phase for low-mass
  binaries with similar masses of the two stars, most of the mass of
  the envelope of the evolved star is lost without a significant
  spiral-in.  This suggestion is supported by the fact that the
  original reasoning for spiral-in (drag forces in the envelope) is
  not applicable here, because the envelope can easily been spun up to
  corotate with the binary.
  
  In the absence of a physical picture for the removal of the
  envelope, we introduce a simple parameter $\gamma = (\Delta J/\Delta
  M_{\rm tot}) \; (M_{\rm tot}/J)$ to describe the loss of the angular
  momentum of the system as in the early work of \citet{pz67}.  Our
  analysis of the observed parameters for all observed double helium
  white dwarfs shows that the material of the envelope of the giant is
  expelled with 1.4 to 1.7 times the specific angular momentum of the
  initial binary. The details of this kind of mass transfer should be
  investigated using 3D gas-dynamical calculations, which are becoming
  available \citep[e.g.][]{bbk+98a}, but are not yet accurate enough
  to make predictions.

\item With some well-constrained assumptions for the masses of the
  unseen companions in the other 8 double white dwarfs, we find
  similar results as for the double white dwarfs with two helium
  components.
\end{enumerate}

\begin{acknowledgements}
  We are indebted to A. Fedorova for making some trial computations of
  stellar evolution upon our request and to the referee for very
  useful remarks.  LRY acknowledges warm hospitality of the
  Astronomical Institute ``Anton Pannekoek''. This work was supported
  by NWO Spinoza grant 08-0 to E.~P.~J.~van den Heuvel, RFBR grant
  99-02-16037, Russian Federal Program `Astronomy' and by NASA through
  Hubble Fellowship grant HF-01112.01-98A awarded (to SPZ) by the
  Space Telescope Science Institute, which is operated by the
  Association of Universities for Research in Astronomy, Inc., for
  NASA under contract NAS\, 5-26555.
\end{acknowledgements}

\bibliography{journals,binaries}
\bibliographystyle{NBaa}

\end{document}